\documentclass[11pt]{article}
\usepackage{graphicx}

\newcommand{\text}{\rm}

\begin{document}

\title{\textbf{Gribov horizon in the presence of dynamical mass generation in
Euclidean Yang-Mills theories in the Landau gauge }}
\author{R.F. Sobreiro, S.P. Sorella\thanks{%
sobreiro@dft.if.uerj.br, sorella@uerj.br} \\
{\small {\textit{UERJ - Universidade do Estado do Rio de Janeiro,}}} \\
{\small {\textit{\ Rua S\~{a}o Francisco Xavier 524, 20550-013 Maracan\~{a},
}}} {\small {\textit{Rio de Janeiro, Brazil.}}} \and D. Dudal\thanks{%
Research Assistant of the Fund for scientific Research-Flanders, Belgium.},\
H. Verschelde\thanks{%
david.dudal@ugent.be, henri.verschelde@ugent.be} \\
{\small {\textit{Ghent University }}}\\
{\small {\textit{Department of Mathematical Physics and Astronomy,
Krijgslaan 281-S9, }}}\\
{\small {\textit{B-9000 Gent, Belgium}}}}
\date{}
\maketitle

\begin{abstract}
The infrared behavior of the gluon and ghost propagators is analyzed in
Yang-Mills theories in the presence of dynamical mass generation in the
Landau gauge. By restricting the domain of integration in the path-integral
to the Gribov region $\Omega $, the gauge propagator is found to be
suppressed in the infrared, while the ghost propagator is enhanced.
\end{abstract}

\vfill\newpage\ \makeatother

\renewcommand{\theequation}{\thesection.\arabic{equation}}

\section{Introduction.}

The possibility that gluons might acquire a mass through a dynamical
mechanism is receiving renewed interest in the last few years. Although a
fully gauge invariant framework for the dynamical mass generation in
Yang-Mills theories is not yet available, the number of gauges displaying
this interesting phenomenon is getting considerably large.

A dynamical gluon mass has been introduced in the light-cone gauge \cite{jc}
in order to obtain estimates for the spectrum of the glueballs. It has been
discussed in the Coulomb gauge in \cite{gh}, where the presence of a
nonvanishing condensate $\left\langle A_{i }^{a}A_{i }^{a}\right\rangle$ in
the operator product expansion for the two-point gauge correlation function
has been pointed out. More recently, the condensate $\left\langle A_{\mu
}^{a}A_{\mu }^{a}\right\rangle $ has been investigated in the Landau gauge
in \cite{z,b}, where it has been proven to account for the discrepancy
observed in the two- and three-point correlation functions between the
perturbative theory and the lattice results. A renormalizable effective
potential for the condensate $\left\langle A_{\mu }^{a}A_{\mu
}^{a}\right\rangle $ in pure Yang-Mills theory in the Landau gauge has been
constructed and evaluated in analytic form up to two-loop order in \cite
{v,v1}. This result shows that the vacuum of Yang-Mills theory favors the
formation of a nonvanishing condensate $\left\langle A_{\mu }^{a}A_{\mu
}^{a}\right\rangle $, which lowers the vacuum energy and provides a
dynamical gluon mass, which turns out to be of the order of $\approx 500MeV$%
. The inclusion of massless quarks has been worked out in \cite{jg}. We
remind here that lattice simulations of the gluon propagator in the Landau
gauge have reported a gluon mass $m\approx 600MeV$ \cite{lg}. Concerning
other gauges, the occurrence of the condensate $\left\langle A_{\mu
}^{a}A_{\mu }^{a}\right\rangle $ and of the related dynamical gluon mass has
been established in the linear covariant gauges in \cite{lin,lin1}. These
results can be generalized to a class of nonlinear covariant gauges. Here,
the mixed gluon-ghost condensate $\left\langle \frac{1}{2}A_{\mu }^{a}A_{\mu
}^{a}+\xi \overline{c}^{a}c^{a}\right\rangle $ has to be considered \cite{m}%
, with $\xi$ the gauge parameter. A renormalizable effective potential for
this condensate has been obtained in the Curci-Ferrari \cite{cf} and Maximal
Abelian gauges \cite{mag}, resulting in a dynamical mass generation. In the
latter case, lattice simulations \cite{maglatt} had already given evidences
of a nonvanishing mass for the off-diagonal gluons. Moreover, a gluon mass
has been reported in lattice simulations in the Laplacian gauge \cite{lapl}.
Also, it is part of the so-called Kugo-Ojima criterion for color confinement
\cite{ko} and, as discussed in \cite{jf}, it proves to be useful in order to
account for the data obtained on the radiative decays of heavy quarkonia
systems.

In this work we pursue the study of the dynamical mass generation
in Euclidean Yang-Mills theory in the Landau gauge. We attempt at
incorporating the nonperturbative effects related to the Gribov
horizon \cite{g}, the aim being that of investigating the infrared
behavior of the gluon and ghost propagators in presence of the
dynamical mass generation. These propagators have been studied to
a great extent by several groups through lattice simulations
\cite{lg,latt1,latt2,latt3,latt4} in the Landau gauge, which have
confirmed that the gluon propagator is suppressed in the infrared
region while the ghost propagator is enhanced, being in fact more
singular than the perturbative behavior $\approx 1/k^{2}$. Such
behavior of the gluon and ghost propagators was already found by
Gribov in \cite{g}, where it arises as a consequence of the
restriction of the domain of integration in the path-integral to
the region $\Omega $ whose boundary $\partial \Omega $ is the
first Gribov horizon, where the first vanishing eigenvalue of the
Faddeev-Popov operator, $-\partial _{\mu }\left( \partial _{\mu
}\delta ^{ab}+gf^{acb}A_{\mu }^{c}\right) $, appears. This
restriction is necessary due to the existence of the Gribov
copies, which imply that the Landau condition, $\partial _{\mu
}A_{\mu }^{a}=0$, does not uniquely fix the gauge. The infrared
suppression of the gluon propagator and the enhancement of the
ghost propagator have also been derived in \cite{z1}, where the
restriction to the region $\Omega $ has been implemented by a
Boltzmann factor through the introduction of a horizon function.
Recently, the authors of \cite{sd,bl,z2,z3} have analyzed the
behavior of the gluon and ghost propagators in the Landau gauge
within the Schwinger-Dyson framework, also obtaining that the
gluon propagator is suppressed while the ghost propagator is
enhanced.

Concerning now the gluon and ghost propagators in the presence of a
dynamical mass generation, we shall proceed by following Gribov's original
suggestion, which amounts to implement the restriction to $\Omega $ as a
no-pole condition for the two-point ghost function \cite{g}. We shall be
able to show that the gluon and ghost propagators are suppressed and
enhanced, respectively, and this in the presence of a dynamical gluon mass.
This behavior is in agreement with that found in \cite{g,z1,sd,bl,z2,z3}.

This work is organized as follows. In Sect.2 we briefly review the
properties of the Lagrangian accounting for the dynamical gluon mass
generation in the Landau gauge. In Sect.3 we implement the restriction of
the domain of integration in the path-integral to the region $\Omega $. The
ensuing modifications of the gauge propagator due to both the Gribov horizon
and dynamical gluon mass are worked out. Sect.4 is devoted to the analysis
of the infrared behavior of the ghost propagator. Some further remarks are
collected in the Conclusion.

\section{Dynamical mass generation in the Landau gauge.}

The dynamical mass generation in the Landau gauge is described by the
following action \cite{v}
\begin{equation}
S(A,\sigma )=S_{YM}+S_{GF+FP}+S_{\sigma }\;,  \label{eq1}
\end{equation}
where $S_{YM}$, $S_{GF+FP}$ are the Yang-Mills and the gauge fixing terms
\begin{equation}
S_{YM}=\frac{1}{4}\int d^{4}xF_{\mu \nu }^{a}F_{\mu \nu }^{a}\;,  \label{eq2}
\end{equation}
\begin{equation}
S_{GF+FP}=\int d^{4}x\left( b^{a}\partial _{\mu }A_{\mu }^{a}+\overline{c}%
^{a}\partial _{\mu }D_{\mu }^{ab}c^{b}\right) \;,  \label{eq3}
\end{equation}
with $b^{a}$ being the Lagrange multiplier enforcing the Landau gauge
condition, $\partial _{\mu }A_{\mu }^{a}=0$, and $\overline{c}^{a},$ $c^{a}$
denoting the Faddeev-Popov ghosts. The color index $a$ refers to the adjoint
representation of the gauge group $SU(N)$. The term $S_{\sigma }$ in eq.$%
\left( \ref{eq1}\right) $ contains the auxiliary scalar field $\sigma $ and
reads

\begin{equation}
S_{\sigma }=\int d^{4}x\left( \frac{\sigma ^{2}}{2g^{2}\zeta }+\frac{1}{2}%
\frac{\sigma }{g\zeta }A_{\mu }^{a}A_{\mu }^{a}+\frac{1}{8\zeta }\left(
A_{\mu }^{a}A_{\mu }^{a}\right) ^{2}\;\right) .  \label{eq4}
\end{equation}
The introduction of the auxiliary field $\sigma $ allows to study the
condensation of the local operator $A_{\mu }^{a}A_{\mu }^{a}$. In fact, as
shown in \cite{v}, the following relation holds
\begin{equation}
\left\langle \sigma \right\rangle =-\frac{g}{2}\left\langle A_{\mu
}^{a}A_{\mu }^{a}\right\rangle \;.  \label{eq5}
\end{equation}
The dimensionless parameter $\zeta $ in expression $\left( \ref{eq4}\right) $
is needed to account for the ultraviolet divergences present in the vacuum
correlation function $\left\langle A^{2}(x)A^{2}(y)\right\rangle $. For the
details of the renormalizability properties of the local operator $A_{\mu
}^{a}A_{\mu }^{a}$ in the Landau gauge we refer to \cite{a2l,a2ll}.
Expression $\left( \ref{eq1}\right) $ is left invariant by the following
BRST transformations
\begin{eqnarray}
sA_{\mu }^{a} &=&-D_{\mu }^{ab}c^{b}=-\left( \partial _{\mu
}c^{a}+gf^{abc}A_{\mu }^{b}c^{c}\right) \;,  \nonumber \\
sc^{a} &=&\frac{1}{2}gf^{abc}c^{b}c^{c}\;,  \nonumber \\
s\overline{c}^{a} &=&b^{a}\;,  \nonumber \\
sb^{a} &=&0\;,  \nonumber \\
s\sigma &=&gA_{\mu }^{a}\partial _{\mu }c^{a}\;,\;  \label{s}
\end{eqnarray}
and
\begin{equation}
sS(A,\sigma )=0\;.  \label{eq6}
\end{equation}
Notice that, from the relation
\begin{equation}
A_{\mu }^{a}\partial _{\mu }c^{a}=-\frac{1}{2}s\left( A_{\mu }^{a}A_{\mu
}^{a}\right) \;,  \label{eq7}
\end{equation}
it follows that the BRST\ operator is nilpotent. The action $S(A,\sigma )$
is the starting point for constructing a renormalizable effective potential $%
V(\sigma )$ for the auxiliary field $\sigma $, obeying the renormalization
group equations. The output of the higher loop computations done in \cite
{v,jg} shows that the minimum of $V(\sigma )$ occurs for a nonvanishing
vacuum expectation value of the auxiliary field, \textit{i.e. }$\left\langle
\sigma \right\rangle \neq 0$. In particular, from expression $\left( \ref
{eq1}\right) $, the first order induced dynamical gluon mass is found to be
\begin{equation}
m^{2}=\frac{g\left\langle \sigma \right\rangle }{\zeta _{0}}\;,  \label{eq8}
\end{equation}
where $\zeta _{0}$ is the first contribution to the parameter $\zeta $ \cite
{v}, given by
\begin{eqnarray}
\zeta &=&\frac{\zeta _{0}}{g^{2}}+\zeta _{1}+O(g^{2})\;,  \nonumber \\
\zeta _{0} &=&\frac{9}{13}\frac{\left( N^{2}-1\right) }{N}\;.  \label{eq9}
\end{eqnarray}
We remind here that, in the Landau gauge, the Faddeev-Popov ghosts $%
\overline{c}^{a},$ $c^{a}$ remain massless, due to the absence of mixing
between the composite operators $A^{a}_{\mu}A^{a}_{\mu}$ and $\overline{c}%
^{a} c^{a}$. This stems from additional Ward identities present in the
Landau \cite{a2ll} and in the covariant linear gauges \cite{lin}, which
forbid the appearance of the term $\overline{c}^{a} c^{a}$.

\section{Infrared behavior of the gluon propagator.}

\subsection{Restriction to the region $\Omega $.}

In the previous section we have reviewed the properties of the action $%
S(A,\sigma )$ which accounts for the dynamical mass generation. However, it
is worth underlining that the action $S(A,\sigma )$ leads to a partition
function
\begin{equation}
\mathcal{Z}=\mathcal{N}\int DAD\sigma \;\delta (\partial
A^{a})\det \left( -\partial _{\mu }\left( \partial _{\mu }\delta
^{ab}+gf^{acb}A_{\mu }^{c}\right) \right) e^{-\left(
S_{YM}+S_{\sigma }\right) }\;,  \label{eq10}
\end{equation}
which is still plagued by the Gribov copies, which affect the Landau gauge$.$
It might be useful to notice here that the action $\left( S_{YM}+S_{\sigma
}\right) $ is left invariant by the local gauge transformations

\begin{eqnarray}
\delta A_{\mu }^{a} &=&-D_{\mu }^{ab}\omega ^{b}\;,\;\;\;\;\;\;  \label{eqqa}
\\
\delta \sigma &=&gA_{\mu }^{a}\partial _{\mu }\omega ^{a}\;,\;  \nonumber
\end{eqnarray}
\begin{equation}
\delta \left( S_{YM}+S_{\sigma }\right) =0\;.  \label{eqqb}
\end{equation}
As a consequence of the existence of Gribov copies, the domain of
integration in the path-integral should be restricted further. We shall
follow here Gribov's proposal to restrict the domain of integration to the
region $\Omega $ \cite{g}. Expression $\left( \ref{eq10}\right) $ is thus
replaced by
\begin{equation}
\mathcal{Z}=\mathcal{N}\int DAD\sigma \;\delta (\partial
A^{a})\det \left( -\partial _{\mu }\left( \partial _{\mu }\delta
^{ab}+gf^{acb}A_{\mu}^{c}\right) \right) e^{-\left( S_{YM}+S_{\sigma }\right) }\;\mathcal{V}%
(\Omega )\;,  \label{eq11}
\end{equation}
where $\mathcal{V}(\Omega )$ implements the restriction to $\Omega
$. The factor $\mathcal{V}(\Omega )$ can be accommodated for by
requiring that the two-point connected ghost function
$\mathcal{G}(k,A)$ has no poles for a given nonvanishing value of
the momentum $k$ \cite{g}. This condition can be understood by
recalling that the region $\Omega $ is defined as the set of all
transverse gauge connections $\left\{ A_{\mu }^{a}\right\} $,
$\partial _{\mu }A_{\mu }^{a}=0,$ for which the Faddeev-Popov
operator is positive definite, \textit{i.e.} $-\partial _{\mu
}\left( \partial _{\mu }\delta ^{ab}+gf^{acb}A_{\mu }^{c}\right)
>0$. This implies that the inverse of the Faddeev-Popov operator
$\left[ -\partial _{\mu }\left( \partial _{\mu
}\delta ^{ab}+gf^{acb}A_{\mu }^{c}\right) \right] ^{-1}$, and thus $\mathcal{%
G}(k,A)$, can become large only when approaching the horizon $\partial
\Omega $, which corresponds in fact to $k=0$ \cite{g}. The quantity $%
\mathcal{G}(k,A)$ can be evaluated order by order in perturbation theory.
Repeating the same calculation of \cite{g}, we find that, up to the second
order

\begin{equation}
\mathcal{G}(k,A)\approx \frac{1}{k^{2}}\frac{1}{1-\rho (k,A)}\;,
\label{eq12}
\end{equation}
with $\rho (k,A)$ given by
\begin{equation}
\rho (k,A)=\frac{g^{2}}{3}\frac{N}{N^{2}-1}\frac{1}{V}\frac{k_{\mu }k_{\nu }%
}{k^{2}}\sum_{q}\frac{1}{\left( k-q\right) ^{2}}\left( A_{\lambda
}^{a}(q)A_{\lambda }^{a}(-q)\right) \left( \delta _{\mu \nu }-\frac{q_{\mu
}q_{\nu }}{q^{2}}\right) \;,  \label{eq13}
\end{equation}
and $V$ being the space-time volume. According to \cite{g}, the no-pole
condition for $\mathcal{G}(k,A)$ reads
\begin{eqnarray}
\rho (0,A) &<&1\;,  \nonumber \\
\rho (0,A) &=&\frac{g^{2}}{4}\frac{N}{N^{2}-1}\frac{1}{V}\sum_{q}\frac{1}{%
q^{2}}\left( A_{\lambda }^{a}(q)A_{\lambda }^{a}(-q)\right) \;.  \label{eq14}
\end{eqnarray}
Therefore, for the factor $\mathcal{V}(\Omega )$ in eq.$\left( \ref{eq11}%
\right) $ we have
\begin{equation}
\mathcal{V}(\Omega )=\theta (1-\rho (0,A))\;,  \label{eq15}
\end{equation}
where $\theta (x)$ stands for the step function\footnote{$\theta (x)=1$ if $%
x>0$, $\theta (x)=0$ if $x<0$.}.\

\subsection{The gauge propagator.}

In order to discuss the gauge propagator, it is sufficient to retain only
the quadratic terms in expression $\left( \ref{eq11}\right) $ which
contribute to the two-point correlation function $\left\langle A_{\mu
}^{a}(k)A_{\nu }^{b}(-k)\right\rangle $. Expanding around the nonvanishing
vacuum expectation value of the auxiliary field, $\left\langle \sigma
\right\rangle \neq 0$, and making use of the integral representation for the
step function
\begin{equation}
\theta (1-\rho (0,A))=\int_{-i\infty +\varepsilon }^{i\infty +\varepsilon }%
\frac{d\eta }{2\pi i\eta }e^{\eta (1-\rho (0,A))}\;,  \label{eq16}
\end{equation}
we get
\begin{eqnarray}
\mathcal{Z}_{\mathrm{quadr}} &=&\mathcal{N}\int DA\frac{d\eta }{2\pi i\eta }%
e^{\eta (1-\rho (0,A))}\delta (\partial A^{a})e^{-\left( \frac{1}{4}\int
d^{4}x(\left( \partial _{\mu }A_{\nu }^{a}-\partial _{\mu }A_{\nu
}^{a}\right) ^{2}+\frac{1}{2}m^{2}\int d^{4}x\left( A_{\mu }^{a}A_{\mu
}^{a}\right) \right) }\;\;  \nonumber \\
&=&\mathcal{N}\int DA\frac{d\eta }{2\pi i\eta }e^{\eta }e^{-\frac{1}{2}%
\sum_{q}A_{\mu }^{a}(q)\mathcal{Q}_{\mu \nu }^{ab}A_{\nu }^{b}(-q)}\;,
\label{eq17}
\end{eqnarray}
with
\begin{equation}
\mathcal{Q}_{\mu \nu }^{ab}=\left( \left( q^{2}+m^{2}\right) \delta _{\mu
\nu }+\left( \frac{1}{\alpha }-1\right) q_{\mu }q_{\nu }+\frac{\eta Ng^{2}}{%
N^{2}-1}\frac{1}{2V}\frac{1}{q^{2}}\delta _{\mu \nu }\right) \delta ^{ab}\;,
\label{eq18}
\end{equation}
where the limit $\alpha \rightarrow 0$ has to be taken at the end in order
to recover the Landau gauge. Integrating over the gauge field, one has
\begin{equation}
\mathcal{Z}_{\mathrm{quadr}}=\mathcal{N}\int \frac{d\eta }{2\pi i\eta }%
e^{\eta }\left( \det \mathcal{Q}_{\mu \nu }^{ab}\right) ^{-\frac{1}{2}}=%
\mathcal{N}\int \frac{d\eta }{2\pi i}e^{f(\eta )}\;,  \label{eq19}
\end{equation}
where $f(\eta )$ is given by
\begin{equation}
f(\eta )=\eta -\log \eta -\frac{3}{2}(N^{2}-1)\sum_{q}\log \left(
q^{2}+m^{2}+\frac{\eta Ng^{2}}{N^{2}-1}\frac{1}{2V}\frac{1}{q^{2}}\right) \;.
\label{eq20}
\end{equation}
Following \cite{g}, the expression $\left( \mathrm{{\ref{eq19}}}\right) $
can be now evaluated at the saddle point, namely
\begin{equation}
\mathcal{Z}_{\mathrm{quadr}}\approx e^{f(\eta _{0})}\;,  \label{eq21}
\end{equation}
where $\eta _{0}$ is determined by the minimum condition
\begin{equation}
1-\frac{1}{\eta _{0}}-\frac{3}{4}\frac{Ng^{2}}{V}\sum_{q}\frac{1}{\left(
q^{4}+m^{2}q^{2}+\frac{\eta _{0}Ng^{2}}{N^{2}-1}\frac{1}{2V}\right) }=0\;.
\label{eq22}
\end{equation}
Taking the thermodynamic limit, $V\rightarrow \infty $, and setting \cite{g}
\begin{equation}
\gamma ^{4}=\frac{\eta _{0}Ng^{2}}{N^{2}-1}\frac{1}{2V}\;\;,\;\;\;\;V%
\rightarrow \infty \;,  \label{eq23}
\end{equation}
we get the gap equation
\begin{equation}
\frac{3}{4}Ng^{2}\int \frac{d^{4}q}{\left( 2\pi \right) ^{4}}\frac{1}{%
q^{4}+m^{2}q^{2}+\gamma ^{4}}=1\;,  \label{eq24}
\end{equation}
where the term $1/\eta _{0}$ in $\left( \mathrm{{\ref{eq22}}}\right) $ has
been neglected in the thermodynamic limit. The gap equation $\left( \mathrm{{%
\ref{eq24}}}\right) $ defines the parameter $\gamma $. Notice that the
dynamical mass $m$ appears explicitly in eq.$\left( \mathrm{{\ref{eq24}}}%
\right) $. Moreover, $\left( \mathrm{{\ref{eq24}}}\right) $ reduces to the
original gap equation of \cite{g,z1} for $m=0$. To obtain the gauge
propagator, we can now go back to the expression for $\mathcal{Z}_{\mathrm{%
quadr}}$ which, after substituting the saddle point value $\eta =\eta _{0}$,
becomes
\begin{equation}
\mathcal{Z}_{\mathrm{quadr}}=\mathcal{N}\int DAe^{-\frac{1}{2}\sum_{q}A_{\mu
}^{a}(q)\mathcal{Q}_{\mu \nu }^{ab}A_{\nu }^{b}(-q)}\;,  \label{eq25}
\end{equation}
with
\begin{equation}
\mathcal{Q}_{\mu \nu }^{ab}=\left( \left( q^{2}+m^{2}+\frac{\gamma ^{4}}{%
q^{2}}\right) \delta _{\mu \nu }+\left( \frac{1}{\alpha }-1\right) q_{\mu
}q_{\nu }\right) \delta ^{ab}\;.  \label{eq26}
\end{equation}
Computing the inverse of $\mathcal{Q}_{\mu \nu }^{ab}$ and taking the limit $%
\alpha \rightarrow 0$, we get the gauge propagator in the presence of the
dynamical gluon mass $m$, \textit{i.e.}
\begin{equation}
\left\langle A_{\mu }^{a}(q)A_{\nu }^{b}(-q)\right\rangle =\delta ^{ab}\frac{%
q^{2}}{q^{4}+m^{2}q^{2}+\gamma ^{4}}\left( \delta _{\mu \nu }-\frac{q_{\mu
}q_{\nu }}{q^{2}}\right) \;.  \label{eq27}
\end{equation}
Notice that, the presence of the mass $m$ in eq.$\left( \mathrm{{\ref{eq27}}}%
\right) $ enforces the infrared suppression of the gluon
propagator.

\section{The infrared behavior of the ghost propagator.}

Let us discuss now the infrared behavior of the ghost propagator, given by
eq.$\left( \ref{eq12}\right) $ upon contraction of the gauge fields, namely

\begin{equation}
\mathcal{G}\approx \frac{1}{k^{2}}\frac{1}{1-\rho (k)}\;,  \label{eq28}
\end{equation}
with
\begin{eqnarray}
\rho (k) &=&\frac{g^{2}}{3}\frac{N}{N^{2}-1}\frac{k_{\mu }k_{\nu }}{k^{2}}%
\int \frac{d^{4}q}{\left( 2\pi \right) ^{4}}\frac{1}{\left( k-q\right) ^{2}}%
\left\langle A_{\lambda }^{a}(q)A_{\lambda }^{a}(-q)\right\rangle \left(
\delta _{\mu \nu }-\frac{q_{\mu }q_{\nu }}{q^{2}}\right) \;  \nonumber \\
&=&g^{2}N\frac{k_{\mu }k_{\nu }}{k^{2}}\int \frac{d^{4}q}{\left( 2\pi
\right) ^{4}}\frac{1}{\left( k-q\right) ^{2}}\frac{q^{2}}{%
q^{4}+m^{2}q^{2}+\gamma ^{4}}\left( \delta _{\mu \nu }-\frac{q_{\mu }q_{\nu }%
}{q^{2}}\right) \;.  \nonumber \\
&&  \label{eq29}
\end{eqnarray}
From the gap equation $\left( \mathrm{{\ref{eq24}}}\right) $, it follows
\begin{equation}
Ng^{2}\int \frac{d^{4}q}{\left( 2\pi \right) ^{4}}\frac{1}{%
q^{4}+m^{2}q^{2}+\gamma ^{4}}\left( \delta _{\mu \nu }-\frac{q_{\mu }q_{\nu }%
}{q^{2}}\right) =\delta _{\mu \nu }\;,  \label{eq30}
\end{equation}
so that
\begin{eqnarray}
1-\rho (k) &=&Ng^{2}\frac{k_{\mu }k_{\nu }}{k^{2}}\int \frac{d^{4}q}{\left(
2\pi \right) ^{4}}\frac{k^{2}-2qk}{\left( k-q\right) ^{2}}\frac{1}{%
q^{4}+m^{2}q^{2}+\gamma ^{4}}\left( \delta _{\mu \nu }-\frac{q_{\mu }q_{\nu }%
}{q^{2}}\right) \;.  \nonumber \\
&&  \label{eq31}
\end{eqnarray}
Notice that the integral in the right hand side of eq.$\left( \mathrm{{\ref
{eq31}}}\right) $ is convergent and nonsingular at $k=0$. Therefore, for $%
k\approx 0$,
\begin{equation}
\left( 1-\rho (k)\right) _{k\approx 0}\approx \frac{3Ng^{2}\mathcal{J}}{4}%
k^{2}\;,  \label{eq32}
\end{equation}
where $\mathcal{J}$ stands for the value of the integral
\begin{equation}
\mathcal{J=}\int \frac{d^{4}q}{\left( 2\pi \right) ^{4}}\frac{1}{%
q^{2}(q^{4}+m^{2}q^{2}+\gamma ^{4})}\;.  \label{eq33}
\end{equation}
Finally, for the ghost propagator we get
\begin{equation}
\mathcal{G}_{k\approx 0}\approx \frac{4}{3Ng^{2}\mathcal{J}}\frac{1}{k^{4}}%
\;,  \label{eq34}
\end{equation}
exhibiting the characteristic infrared enhancement which, thanks to the gap
equation $\left( \mathrm{{\ref{eq24}}}\right) $, turns out to hold also in
the presence of the dynamical mass generation.

\section{Conclusion.}

In this letter we have analyzed the infrared behavior of the gluon and ghost
propagators in the presence of dynamical mass generation in the Landau
gauge. The restriction of the domain of integration to the Gribov region $%
\Omega $ has been implemented by repeating Gribov's procedure \cite{g},
which amounts to impose a no-pole condition for the two-point ghost
function. The output of our analysis is summarized by equations $\left(
\mathrm{{\ref{eq24}}}\right) $, $\left( \mathrm{{\ref{eq27}}}\right) $, $%
\left( \mathrm{{\ref{eq34}}}\right) $. Expression $\left( \mathrm{{\ref{eq24}%
}}\right) $ is the gap equation which defines the parameter $\gamma $.
Notice now that the dynamical mass $m$ enters explicitly the gap equation
for $\gamma $. Equation $\left( \mathrm{{\ref{eq27}}}\right) $ yields the
gauge propagator, which exhibits the infrared suppression. Finally, equation
$\left( \mathrm{{\ref{eq34}}}\right) $ establishes the enhancement of the
ghost propagator. This behavior of the gluon and ghost propagators is in
agreement with that found in \cite{g,z1,sd,bl,z2,z3}. Also, lattice
simulations \cite{lg,latt1,latt2,latt3,latt4} have provided confirmations of
the infrared suppression of the gluon propagator and of the ghost
enhancement, in the Landau gauge.

Concerning now the Gribov region $\Omega $, it is known that it is not free
from Gribov copies \cite{c1,c2,c3}. The uniqueness of the gauge condition
should be ensured by restricting the domain of integration to a smaller
region in field space, known as the fundamental modular region. However,
this is a difficult task. Nevertheless, the restriction to the Gribov region
$\Omega $ captures nontrivial nonperturbative aspects of the infrared
behavior of the theory, as expressed by the suppression and the enhancement
of the gluon and ghost propagators. Recently, it has been argued in \cite{z3}
that the additional copies present in the Gribov region $\Omega $ might have
no influence on the expectation values.

Although being outside of the aim of the present letter, we remark that the
gap equation $\left( \mathrm{{\ref{eq24}}}\right) $ can be also derived by
using as starting point the local renormalizable action implementing the
Gribov horizon, proposed in \cite{z1} by Zwanziger. It turns out in fact
that the local operator $A_{\mu }^{a}A_{\mu }^{a}$ can be added to the
Zwanziger action without spoiling its renormalizability \cite{prep}. This
will allow to study the condensation of the operator $A_{\mu }^{a}A_{\mu
}^{a}$ when the restriction to the horizon is taken into account. In this
case, the combination of the algebraic BRST technique with the local
composite operator formalism, see e.g. \cite{v,lin1,a2ll}, should make
possible to include the renormalization effects on the gluon and ghost
propagators and to see how well these compare with the available lattice
results.

\section*{Acknowledgments.}

The Conselho Nacional de Desenvolvimento Cient\'{\i}fico e Tecnol\'{o}gico
(CNPq-Brazil), the SR2-UERJ and the Coordena{\c{c}}{\~{a}}o de Aperfei{\c{c}}%
oamento de Pessoal de N{\'\i}vel Superior (CAPES) are gratefully
acknowledged for financial support. D. Dudal would like to thank the
Theoretical Physics Department of the UERJ for the kind hospitality.


\begin{thebibliography}{99}
\bibitem{jc}  J.~M.~Cornwall, Phys.\ Rev.\ D \textbf{26} (1982) 1453.

\bibitem{gh}  J.~Greensite and M.~B.~Halpern, Nucl.\ Phys.\ B \textbf{271}
(1986) 379.

\bibitem{z}  F.~V.~Gubarev, L.~Stodolsky and V.~I.~Zakharov, Phys.\ Rev.\
Lett.\ \textbf{86} (2001) 2220;

F.~V.~Gubarev and V.~I.~Zakharov, Phys.\ Lett.\ B \textbf{501} (2001).

\bibitem{b}  P.~Boucaud, A.~Le Yaouanc, J.~P.~Leroy, J.~Micheli, O.~P\`{e}ne
and J.~Rodriguez-Quintero, Phys.\ Lett.\ B \textbf{493} (2000) 315;

P.~Boucaud, A.~Le Yaouanc, J.~P.~Leroy, J.~Micheli, O.~P\`{e}ne and
J.~Rodriguez-Quintero, Phys.\ Rev.\ D \textbf{63} (2001) 114003;

P.~Boucaud, J.~P.~Leroy, A.~Le Yaouanc, J.~Micheli, O.~P\`{e}ne, F.~De Soto,
A.~Donini, H.~Moutarde and J. Rodr{\'{i}}guez-Quintero, Phys.\ Rev.\ D
\textbf{66} (2002) 034504.

\bibitem{v}  H.~Verschelde, K.~Knecht, K.~Van Acoleyen and M.~Vanderkelen,
Phys.\ Lett.\ B \textbf{516} (2001) 307.

\bibitem{v1}  D.~Dudal, H.~Verschelde, R.~E.~Browne and J.~A.~Gracey, Phys.\
Lett.\ \textbf{B} 562 (2003) 87.

\bibitem{jg}  R.~E.~Browne and J.~A.~Gracey, JHEP \textbf{0311} (2003) 029.

\bibitem{lg}  K.~Langfeld, H.~Reinhardt and J.~Gattnar, Nucl.\ Phys.\ B
\textbf{621} (2002) 131.

\bibitem{lin}  D.~Dudal, H.~Verschelde, V.~E.~R.~Lemes, M.~S.~Sarandy,
R.~F.~Sobreiro, S.~P.~Sorella and J.~A.~Gracey, Phys.\ Lett.\ B \textbf{574}
(2003) 325.

\bibitem{lin1}  D.~Dudal, H.~Verschelde, J.~A.~Gracey, V.~E.~R.~Lemes,
M.~S.~Sarandy, R.~F.~Sobreiro and S.~P.~Sorella, JHEP \textbf{01 }(2004) 044.

\bibitem{m}  K.~I.~Kondo, Phys.\ Lett.\ B \textbf{514} (2001) 335;

K.~I.~Kondo, T.~Murakami, T.~Shinohara and T.~Imai, Phys.\ Rev.\ D \textbf{65%
} (2002) 085034.

\bibitem{cf}  D.~Dudal, H.~Verschelde, V.~E.~R.~Lemes, M.~S.~Sarandy,
S.~P.~Sorella and M.~Picariello, Annals Phys.\ \textbf{308} (2003) 62.

\bibitem{mag}  D.~Dudal et al., to appear.

\bibitem{maglatt}  K.~Amemiya and H.~Suganuma, Phys.\ Rev.\ D \textbf{60}
(1999) 114509;

V.~G.~Bornyakov, M.~N.~Chernodub, F.~V.~Gubarev, S.~M.~Morozov and
M.~I.~Polikarpov, Phys.\ Lett.\ B \textbf{559} (2003) 214.

\bibitem{lapl}  C.~Alexandrou, P.~de Forcrand and E.~Follana, Phys.\ Rev.\ D
\textbf{65} (2002) 114508;

C.~Alexandrou, P.~de Forcrand and E.~Follana, Phys.\ Rev.\ D \textbf{65}
(2002) 117502.

\bibitem{ko}  T.~Kugo and I.~Ojima, Prog.\ Theor.\ Phys.\ Suppl.\ \textbf{66}
(1979) 1;

T.~Kugo and I.~Ojima, \textit{Massive Gauge Boson Implies Spontaneous
Breakdown of the Global Gauge Symmetry: Higgs Phenomenon and Quark
Confinement}, Print-79-0268 (Kyoto), KUNS-486, Feb 1979.

\bibitem{jf}  J.~H.~Field, Phys. Rev. D \textbf{66} (2002) 013013.

\bibitem{g}  V.~N.~Gribov, Nucl.\ Phys.\ B \textbf{139} (1978) 1.

\bibitem{latt1}  P.~Marenzoni, G.~Martinelli and N.~Stella, Nucl. Phys. B
\textbf{455} (1995) 339.

\bibitem{latt2}  A.~Cucchieri, Nucl. Phys. B \textbf{508} (1997) 353;

A.~Cucchieri, Phys. Lett. B \textbf{422} (1998) 233;

A.~Cucchieri, Phys. Rev. D \textbf{60} (1999) 034508;

A.~Cucchieri, T.~Mendes and A.~R.~Taurines, Phys. Rev. D \textbf{67} (2003)
091502;

J.~C.~R.~Bloch, A.~Cucchieri, K.~Langfeld and T.~Mendes, hep-lat/0312036.

\bibitem{latt3}  F.~D.~R.~Bonnet, P.~O.~Bowman, D.~B.~Leinweber,
A.~G.~Williams and J.~M.~Zanotti, Phys. Rev. D \textbf{64} (2001) 034501.

\bibitem{latt4}  S.~Furui and H.~Nakajima, hep-lat/0309166;

S.~Furui and H.~Nakajima, hep-lat/0309165;

S.~Furui and H.~Nakajima, hep-lat/0305010.

\bibitem{z1}  D.~Zwanziger, Nucl.\ Phys.\ B \textbf{323} (1989) 513;

D.~Zwanziger, Nucl.\ Phys.\ B \textbf{399} (1993) 477.

\bibitem{sd}  L.~von Smekal, R.~Alkofer and A.~Hauck, Phys. Rev. Lett.
\textbf{79 }(1997) 3591;

L.~von Smekal, A.~Hauck and R.~Alkofer, Annals Phys. \textbf{267 (}1998) 1,
Erratum-ibid. \textbf{269 (}1998) 182;

R.~Alkofer and L.~von Smekal, Phys.\ Rept.\ \textbf{353} (2001) 281;

P.~Watson and R.~Alkofer, Phys.\ Rev.\ Lett. \textbf{86} (2001) 5239.

\bibitem{bl}  D.~Atkinson and J.~C.~R.~Bloch, Phys. Rev. D \textbf{58}
(1998) 094036;

D.~Atkinson and J.~C.~R.~Bloch, Mod. Phys. Lett. A \textbf{13 }(1998) 1055.

\bibitem{z2}  D.~Zwanziger, Phys. Rev. D \textbf{65 }(2002) 094039;

D.~Zwanziger, Phys. Rev. D \textbf{67 }(2003) 105001.

\bibitem{z3}  D.~Zwanziger, Phys. Rev. D \textbf{69} (2004) 016002.

\bibitem{a2l}  J.~A.~Gracey, Phys.\ Lett.\ B \textbf{552} (2003) 101.

\bibitem{a2ll}  D.~Dudal, H.~Verschelde and S.~P.~Sorella, Phys.\ Lett.\ B
\textbf{555} (2003) 126.

\bibitem{c1}  Semenov-Tyan-Shanskii and V.A. Franke, Zapiski Nauchnykh
Seminarov Leningradskogo Otdeleniya Matematicheskogo Instituta im. V.A.
Steklov AN SSSR\}, Vol. \textbf{120} (1982) 159. English translation: New
York: Plenum Press 1986.

\bibitem{c2}  G.~Dell'Antonio and D.~Zwanziger, Commun. Math. Phys. \textbf{%
138} (1991) 291;

G. Dell'Antonio and D.~Zwanziger: Proceedings of the NATO Advanced Research
Workshop on Probabilistic Methods in Quantum Field Theory and Quantum
Gravity, Carg\`{e}se, August 21-27, 1989, Damgaard and Hueffel (eds.),
p.107, New York: Plenum Press.

\bibitem{c3}  P.~van Baal, Nucl. Phys. B \textbf{369} (1992) 259;

P.~van Baal, QCD in a finite volume, in the Boris Ioffe Festschrift, ed. by
M. Shifman, World Scientific. In *Shifman, M. (ed.): At the frontier of
particle physics, vol. 2* 683-760, e-Print Archive: hep-ph/0008206.

\bibitem{prep}  D.~Dudal, H.~Verschelde, R. F. Sobreiro and S.P. Sorella, in
preparation.
\end{thebibliography}
\end{document}